\documentclass[aps,prd,reprint,preprintnumbers]{revtex4-1}
\usepackage{url}
\usepackage{textcase}
\usepackage{bm}
\usepackage{amsmath}
\usepackage{amsfonts}
\usepackage{amssymb}
\usepackage{hyperref}
\usepackage[table]{xcolor}
\usepackage{multirow}
\usepackage{bbm}
\usepackage{graphicx}
\usepackage{tensor}
\usepackage{subfigure}
\usepackage{array}
\usepackage[normalem]{ulem}
\usepackage{float}
\usepackage{fancyhdr}

\usepackage{pdfpages}
\makeatletter
\AtBeginDocument{\let\LS@rot\@undefined}
\makeatother

\usepackage{natbib}
\renewcommand{\Re}{\mathrm{Re}}
\renewcommand{\vec}[1]{\boldsymbol{#1}}

\begin{document}

\title{$\mathcal{N}=2$ supersymmetry and anisotropic scale invariance}
\date{\today}
\author{Daniel K.~Brattan}
\email{danny.brattan@gmail.com}
\affiliation{Interdisciplinary Center for Theoretical Study, University of Science and Technology of China, 96 Jinzhai Road, Hefei, Anhui, 230026 PRC.}

\begin{abstract}
{\ We find a class of scale-anomaly-free $\mathcal{N}=2$ supersymmetric quantum systems with non-vanishing potential terms where space and time scale with distinct exponents. Our results generalise the known case of the supersymmetric inverse square potential to a larger class of scaling symmetries.}
\end{abstract}

\preprint{USTC-ICTS-18-03} 

\maketitle

{\ The breaking of classical symmetries at the quantum level has been a very active area of research since its discovery \cite{PhysRev.177.2426,Bell1969,PhysRevD.34.674,1993AmJPh..61..142H}. A special class of the so-called ``anomalies'' describes the breaking of continuous scale invariance completely or down to an infinite discrete subgroup. Familiar models that suffer from a scale anomaly include a non-relativistic particle in the presence of an inverse square radial potential $\hat{h}_S= p^2/2m - \lambda/r^2$ \cite{Case1950,deAlfaro1976,landau1991quantum,Camblong:2000ec,PhysRevD.68.025006,Hammer:2005sa,Braaten2004,Kaplan:2009kr}, the charged and massless Dirac fermion interacting with a Coulomb potential $\hat{h}_D = \gamma^0 \gamma^j p_j - \lambda/r$ \cite{ovdat2017observing} and a class of one dimensional Lifshitz scalars \cite{Alexandre:2011kr} with $\hat{h}_L = \left(p^2 \right)^N - \lambda/x^{2N}$ \cite{Brattan:2017yzx}.}

{\ To understand the appearance of an anomaly in these systems consider that the classical scaling symmetry of these Hamiltonians permits either no bound states or a continuum; otherwise scale invariance would be broken. For a sufficiently strong, attractive potential our intuition leads us to expect the appearance of at least one negative energy bound state and therefore we expect a continuum. However, an unbounded continuum of normalisable states would imply that the Hamiltonian is non-self-adjoint \cite{Bonneau:1999zq,2015AnHP...16.2367I}. Ensuring that the Hamiltonian is self-adjoint by applying boundary conditions on the elements of the Hilbert space, through the procedure of self-adjoint extension \cite{Gitman:2009era}, must therefore generically break the scale symmetry and introduce an anomaly. One finds precisely the same results by the alternate, and physically pragmatic, method of introducing a cut-off into the system and renormalising \cite{Camblong:2000ec}.}

{\ After the discovery of a scale anomaly in the case of the inverse square potential, interest in $\hat{h}_{S}$ spread to its embedding in a system with $\mathcal{N}=2$ supersymmetry \cite{Akulov1983,FUBINI198417}. This and related models have variously been used to examine exotic forms of supersymmetry breaking \cite{Jevicki:1984xf,Casahorran:1990ck,Imbo:1983cd,Roy:1985qi,Panigrahi:1993zy,Cooper:1994eh,doi:10.1063/1.522710,Falomir:2005yw}, effective theories near black hole horizons \cite{Claus:1998ts,Cadoni:2000gm,Astorino:2002bj} and potentially $M$-theory \cite{Okazaki:2014sga,Okazaki:2015pfa}. Requiring supersymmetry restricts the choice of self-adjoint extension as, in addition to the Hamiltonian being self-adjoint, the supercharges must also be self-adjoint. For $\hat{h}_{S}$ there are subsequently only two choices of self-adjoint extension (see for example \cite{ROY199259}) and the spectrum of bound states is empty. This latter fact follows since supersymmetry permits only positive energy bound states; and there are no normalisable solutions to the energy eigenvalue equation of $\hat{h}_{S}$ with positive energy.}

{\ Subsequently all the physics of the supersymmetric version of $\hat{h}_{S}$ lies in the scattering sector. It is not hard to show, as we shall review in a subsequent section, that the phase shift (equivalently the $S$-matrix on the half line) is a constant independent of the scattering wavelength for the supersymmetric choices of self-adjoint extension. This is exactly what one would expect of a scale invariant theory as there is only one scale in the scattering problem, the wavelength $k$, and thus no way to form a scale invariant ratio.}

{\ Recent work \cite{Brattan:2017yzx,2018arXiv180410213B} has extended the simple class of systems with anomalously broken scale invariance to those where space and time scale with distinct exponents. These Hamiltonians (radial effective Hamiltonians if working in more than one-dimension) have the form
    \begin{eqnarray}
    \label{Eq:ExpandedScaleCovariantHamiltonian}
    \hat{h} &=& \left(p^2\right)^{N} + \sum_{k=1}^{2N} \frac{\lambda_{k}}{r^{k}} d_{r}^{2N-k} \; , \; \; r \in [0,\infty) \; ,
  \end{eqnarray}
where $\lambda_{k}$ are real couplings. Examples of anisotropic scaling symmetry, or ``Lifshitz scaling symmetry'' \cite{Alexandre:2011kr}, can be found at the finite temperature multicritical points of certain materials \cite{PhysRevLett.35.1678,PhysRevB.23.4615} or in strongly correlated electron systems \cite{PhysRevB.69.224415,PhysRevB.69.224416,Ardonne:2003wa}. In particular excitons living on the interfaces of suitable bilayers \cite{2016PhRvB..93w5110S} and ultracold gases with shaken optical lattices \cite{2015NatCo...6E8012P,2015PhRvA..91c3404M,2015PhRvA..91f3634R,2017PhRvB..96h5140W} have quartic dispersion relations ($N=2$) with vanishingly small band gaps. This symmetry may also have applications in particle physics \cite{Alexandre:2011kr}, cosmology \cite{Mukohyama:2010xz} and quantum gravity \cite{PhysRevD.57.971,Kachru:2008yh,Horava:2009if,Horava:2009uw,Gies:2016con}. In fact, the non-interacting mode ($\lambda_{k}=0$) can also appear very generically, for example in non-relativistic systems with spontaneous symmetry breaking \cite{Brauner:2010wm}.}

{\ A natural question poses itself: given a supersymmetric embedding of \eqref{Eq:ExpandedScaleCovariantHamiltonian} is the scale symmetry still anomalous or, like in the case of $\hat{h}_{S}$ \cite{Falomir:2005yw}, is the system scale invariant? These models would be an interesting starting point for tackling field theoretic questions about anisotropic scale invariance. Moreover, the supercharges will be higher derivative and the corresponding systems could be relevant for investigations into polynomial SUSY \cite{Andrianov:1994aj,Andrianov:2003dg,Cannata:2015rva}.} 

{\ The answer to the posed question is given by performing a self-adjoint extension of \eqref{Eq:ExpandedScaleCovariantHamiltonian}. The choices of extension are parameterised by a $(m \times m)$- unitary matrix $(m \leq N)$. This parameter is restricted by imposing supersymmetry and we find that the resultant bound state spectrum is empty (essentially for the same reasons as the $N=1$ case as represented by $\hat{h}_{S}$). Subsequently, there is a subset of supersymmetric choices of self-extension parameter that lead to a scale invariant theory, but a much larger class which do not. In this paper we give the precise condition on the self-adjoint extension parameter to yield a scale invariant theory and discuss how scale invariance is broken in the absence of bound states for the other choices.}

\section{The $\mathcal{N}=2$ SUSY algebra}

{\ A quantum mechanical system is said to have $\mathcal{N}=2$ supersymmetry if there exists two self-adjoint operators $\hat{Q}_{1}$ and $\hat{Q}_{2}$, called the supercharges, such that
  \begin{eqnarray}
   \label{Eq:N=2SUSYdef}
   \left\{ \hat{Q}_{i}, \hat{Q}_{j} \right\} = 2 \delta_{ij} \hat{H}
  \end{eqnarray}
where $\hat{H}$ is a Hamiltonian (see for example \cite{junker1996supersymmetric}); sometimes called the super-Hamiltonian to distinguish it from a regular Hamiltonian. The square of either supercharge equals the super-Hamiltonian and therefore we expect the following identity to be satisfied:
  \begin{eqnarray}
   \left[ \hat{H}, \hat{Q}_{j} \right] = 0 \; .
  \end{eqnarray}
This makes the supercharge a conserved quantity. Importantly, due to a theorem by von Neumann (see for example \cite{reed1980methods}), the square of a self-adjoint operator is also self-adjoint. Hence the super-Hamiltonian is a self-adjoint operator making it consistent with representing the generator of unitary time evolution.}

{\ Consider the differential operators $\hat{Q}_{j}$ defined by:
  \begin{eqnarray}
    \label{Eq:Qjdef}
	\left( \begin{array}{cc} 0 & (-i)^{j-1} \hat{D}_{2N} \ldots \hat{D}_{N+1} \\ (+i)^{j-1} \hat{D}_{N} \ldots \hat{D}_{1} & 0 \end{array} \right) \; , \qquad
  \end{eqnarray}
where
  \begin{eqnarray}
	\label{Eq:SimpleDi}
	\hat{D}_{k}
    &=& i \left( d_{r} - \frac{\Delta_{k} - k + 1}{r} \right) \; , 
  \end{eqnarray}
with $\Delta_{k}$ any set of $2N$ complex numbers. These scale uniformly under $r \mapsto \Lambda r$ and satisfy the anti-commutation relation \eqref{Eq:N=2SUSYdef} which can be re-expressed as:
  \begin{eqnarray}
   \label{Eq:GenericSuperHam}
    \hat{H} = ( \hat{Q}_{1} )^{2} = ( \hat{Q}_{2} )^{2} = \left( \begin{array}{cc} \hat{h} & 0 \\ 0 & \hat{h}' \end{array} \right) \; , \; \;
     \left\{ \hat{Q}_{1}, \hat{Q}_{2} \right\} = 0 \; , \qquad
  \end{eqnarray}
where $\hat{h}$ and $\hat{h}'$ are differential operators of the form \eqref{Eq:ExpandedScaleCovariantHamiltonian}. We avoid the degenerate case $\hat{h} = \hat{h}'$ as in this situation it is very easy to find positive energy bound states \footnote{Given any self-adjoint Hamiltonian $\hat{h}$ of form \eqref{Eq:ExpandedScaleCovariantHamiltonian} with bound states, its square is also of the form \eqref{Eq:ExpandedScaleCovariantHamiltonian} and self-adjoint. This square Hamiltonian has a positive energy spectrum. One can use this to readily generate systems with geometric towers of positive energy bound states by starting with an initial $\hat{h}$ having discrete scale invariance at negative energy \cite{2018arXiv180410213B}.}. The $\Delta_{k}$ are related to the operator $\hat{h}$ by being solutions to $\hat{h} r^{\Delta} = 0$ i.e.~they define the power law behaviour of zero modes.}

{\ Ensuring that the $\hat{Q}_{j}$ of \eqref{Eq:Qjdef} are self-adjoint operators on the Hilbert space will be the prime focus of the subsequent section. This requires a detailed consideration of the boundary conditions at $r=0$ obeyed by wavefunctions in the Hilbert space. For discussions at the level of the SUSY algebra it will be sufficient for us to consider the ``formal adjoints'' which are defined exactly as the adjoints, assuming that all boundary terms can be set to zero. The formal adjoint of $\hat{Q}_{j}$ given in \eqref{Eq:Qjdef} is
  \begin{eqnarray}
   \label{Eq:adjQjdef}
	\left( \begin{array}{cc} 0 & (-i)^{j-1} \hat{D}_{1}^{\dagger} \ldots \hat{D}_{N}^{\dagger} \\  (+i)^{j-1} \hat{D}_{N+1}^{\dagger} \ldots \hat{D}_{2N}^{\dagger} & 0 \end{array} \right) \; , \qquad
  \end{eqnarray}
where
  \begin{eqnarray}
	\label{Eq:SimpleadjDi}
	\hat{D}_{k}^{\dagger}
    &=& i \left( d_{r} + \frac{\Delta_{k}^{*} - k + 1}{r} \right) \; .
  \end{eqnarray}
As the operators $\hat{Q}_{j}$ must at least be formally self-adjoint if they are to be self-adjoint we can see by equating \eqref{Eq:Qjdef} and \eqref{Eq:adjQjdef} that the $\Delta_{i}$ must obey some constraints.}

{\ We shall only consider systems with power laws that do not have real parts equal to $N-1/2$. Such power law behaviours are the origin of discrete scale invariance at negative energies \cite{2018arXiv180410213B}. The appearance of the corresponding infinite set of negative energy bound states is independent of the boundary condition and incompatible with supersymmetry. With such a constraint it follows from \eqref{Eq:SimpleDi} and \eqref{Eq:SimpleadjDi} that: 
  \begin{eqnarray}
   \label{Eq:PowerLawConstraints}
    &\;& \Delta_{2N-j+1} = 2N-1 - \Delta_{j}^{*} \;  , \; \; 1 \leq j \leq N \; . \qquad
  \end{eqnarray}
Finally, if the operators $\hat{h}$ and $\hat{h}'$ are to have real couplings then $\Delta_{j}^{*}$ must belong to the set of power laws whenever $\Delta_{j}$ does; although the supercharges themselves need not be real.}

{\ Given the constraint \eqref{Eq:PowerLawConstraints}, the operators $\hat{H}$, $\hat{h}$ and $\hat{h}'$ are all equal to their formal adjoints. Conversely, it is not difficult to find, given any Hamiltonian $\hat{h}$ of the form \eqref{Eq:ExpandedScaleCovariantHamiltonian} with power laws that meet our constraints, a $\hat{Q}_{j}$ satisfying \eqref{Eq:GenericSuperHam}. However once the power laws of the supplied $\hat{h}$ are determined there is more than one way to label them. This corresponds to swapping pairs of power laws satisfying \eqref{Eq:PowerLawConstraints} between the off-diagonal components of the supercharges $\hat{Q}_{j}$. These distinct supercharges give rise to $2^{N}$ different partner Hamiltonians $\hat{h}'$ corresponding to the original $\hat{h}$.}

\section{Self-adjoint extension}

{\ We now turn to ensuring that the operators $\hat{Q}_{j}$ are self-adjoint on the space where they act. Typically they will not be and we need to perform a self-adjoint extension. When such a self-adjoint extension exists it is parameterised by a unitary matrix of some dimension. To determine if such an extension exists and the size of the matrix we use von Neumann's method \cite{Meetz1964,Bonneau:1999zq,Gitman:2009era,gitman2012self,2015AnHP...16.2367I}. This consists of finding the linearly independent, imaginary eigenvalue, normalisable solutions to
  \begin{eqnarray}
   \label{Eq:ImaginaryEnergyEigenvalueEqn}
   \hat{Q}_{j} \Psi(r) = \pm i \Psi(r) \; , 
  \end{eqnarray}
and counting their number without imposing boundary conditions at $r=0$. Let there be $M_{+}$ solutions with $+i$ and $M_{-}$ solutions with $-i$. If $M_{+} = M_{-} = m \neq 0$ then there is a $U(m)$ self-adjoint extension. If $M_{+} \neq M_{-}$ then no self-adjoint extension exists. Finally if $M_{+}=M_{-}=0$ then the operator is essentially self-adjoint and the wavefunctions upon which $Q_{j}$ act vanish identically in an open region about the origin.}

{\ Let $\Psi(r) = (\psi(r), \tilde{\psi}(r))$ be a generic eigenstate of the operator $\hat{Q}_{j}$ with eigenvalue $\nu \neq 0$. The square of $\nu$ will be the corresponding energy of this eigenstate due to \eqref{Eq:GenericSuperHam}. The eigenvalue equation for $\hat{Q}_{1}$ can be rearranged into the form
  \begin{eqnarray}
   \label{Eq:GenericEigenstateofQplusBosonic}
   \left( \hat{h} - \nu^2 \right) \psi(r) &=& 0 \; , \\
   \label{Eq:GenericEigenstateofQplusFemionic}
    \tilde{\psi}(r) &=& \frac{1}{\nu} \hat{D}_{N} \ldots \hat{D}_{1} \psi(r) \; .
  \end{eqnarray}
The solutions of \eqref{Eq:GenericEigenstateofQplusBosonic} are known analytic functions - the generalised hypergeometric functions (see supplementary note 1). Setting $\nu = \pm i$; we find that for both signs of $i$ there are $N$ decaying solutions to \eqref{Eq:GenericEigenstateofQplusBosonic}.  Normalisability of $\psi(r)$ at $r=0$ however requires that we eliminate by taking linear combinations of these decaying wavefunctions any power law $\Delta_{j}$ from our generic wavefunction with $\mathrm{Re}[\Delta_{j}] \leq -1/2$. This reduces the size of the self-adjoint extension by one for each power law lost. Substituting the generic solution into \eqref{Eq:GenericEigenstateofQplusFemionic} additionally implies that we must remove any power law where $\mathrm{Re}[\Delta_{N+j}] \leq N-1/2$, $1 \leq j \leq N$ so that $\tilde{\psi}(r)$ is normalisable at $r=0$. The result of this process is $m$ linearly independent solutions to \eqref{Eq:ImaginaryEnergyEigenvalueEqn} of both signs and thus a $U(m)$ self-adjoint extension.}

{\ Different self-adjoint extensions correspond to different choices of boundary condition for the wavefunction at $r=0$. The allowed boundary conditions can be determined by taking a generic solution to the equation of motion for $\vec{\Psi}(r)$ and asking what constraints this solution must satisfy so that
  \begin{eqnarray}
    \label{Eq:BoundaryForm}
    \int_{r=0}^{\infty} dr \; \left[ \vec{\Psi}^{\dagger}(r) \hat{Q}_{j} \vec{\Psi}(r) - \left( \hat{Q}_{j} \vec{\Psi}(r) \right)^{\dagger} \vec{\Psi}(r)  \right] \; , \; \; \;
  \end{eqnarray}
vanishes non-trivially \cite{gitman2012self}. The integral \eqref{Eq:BoundaryForm} evaluates to a boundary term which, if we assume decay at infinity, only has contributions from $r=0$.}

{\ At small $r$ every solution to \eqref{Eq:GenericEigenstateofQplusBosonic} with $\nu \neq 0$ has the form
  \begin{eqnarray}
    \label{Eq:Nearr=0BosonExpansion}
    \psi(r) &=& \sum_{j=1}^{N} c_{j} r^{\Delta_{j}} \left( 1 + \mathcal{O}(r^{2N}) \right)\nonumber \\
	       &\;&  + (-i)^{N} \nu \sum_{j=1}^{N} \alpha_{j} \tilde{c}_{j} r^{\Delta_{N+j}} \left( 1 + \mathcal{O}(r^{2N}) \right) \; , \qquad \\
           \left( \alpha_{j} \right)^{-1} &=& \prod_{l=1}^{N} \left( \Delta_{j+N} - \Delta_{l} \right) \; , 
  \end{eqnarray}
where we have assumed that none of the roots differ by $2Nz$, $z \in \mathbbm{Z}$ otherwise we would have to introduce logarithms in this Frobenius solution. The procedure in this degenerate case is given by taking a limit and we shall not concern ourselves with it here. Using \eqref{Eq:GenericEigenstateofQplusFemionic} we can also find the expansion of $\tilde{\psi}(r)$ near $r=0$ to be:
  \begin{eqnarray}
    \label{Eq:Nearr=0FermionExpansion}
    \tilde{\psi}(r) &=& \sum_{j=1}^{N} \tilde{c}_{j} r^{\Delta_{N+j}-N} \left( 1 + \mathcal{O}(r^{2N}) \right) \nonumber \\
	       &\;& + (-i)^N \nu  \sum_{j=1}^{N} \beta_{j} c_{j} r^{\Delta_{j}+N} \left( 1 + \mathcal{O}(r^{2N}) \right)  \; , \qquad \\
    \left( \beta_{j} \right)^{-1} &=& \prod_{l=1}^{N} \left( \Delta_{j} - \Delta_{l+N} + 2 N \right) \; .
  \end{eqnarray}
Imposing normalisability of $\psi(r)$ sets the relevant $c_{j}$ or $\tilde{c}_{j}$ to zero if $\mathrm{Re}[\Delta_{j}] \leq -1/2$. Additionally, imposing normalisability of $\tilde{\psi}(r)$ sets $\tilde{c}_{j} = 0$ if $\mathrm{Re}[\Delta_{j+N}] \leq N-1/2$. Substituting these expressions for $\psi(r)$ and $\tilde{\psi}(r)$ into \eqref{Eq:BoundaryForm} yields a sesquilinear form in terms of the $c_{j}$ and $\tilde{c}_{j}$ (see supplementary note 2) which we are instructed to set to zero. The result can diagonalised in terms of
  \begin{eqnarray}
    c_{j}^{\pm} &=& \frac{1}{\sqrt{2}} \left( \gamma_{j}^{\frac{1}{2}} c_{j} \mp (-i)^{N+1} (\gamma_{j}^{*})^{\frac{1}{2}} \tilde{c}_{N-j+1} \right) \; , 
  \end{eqnarray}
into an expression proportional to
  \begin{eqnarray}
    \label{Eq:BoundaryFormDiagQ1zerocutoff}
    i \left[ | \vec{c}^{+} |^2 - | \vec{c}^{-} |^2 \right] \; ,
  \end{eqnarray}
where $\vec{c}^{\pm}$ are vectors of dimension $m$ and $\gamma_{j}$ is some number dependent on the $\Delta_{l}$. As \eqref{Eq:BoundaryFormDiagQ1zerocutoff} is the difference of the norms of two vectors in $m$-dimensions the general solution which sets it to zero is
    \begin{eqnarray}
   \label{Eq:Q1zerocutoffboundaryconditions}
   \vec{c}^{+} = U_{m} \vec{c}^{-} \; , 
  \end{eqnarray}
where $U_{m}$ is an arbitrary $(m \times m)$-unitary matrix. This matrix is an additional parameter in our model which must be fixed by physical information. Given any system with a Hamiltonian of the form \eqref{Eq:ExpandedScaleCovariantHamiltonian} (satisfying our constraints on the power laws) and boundary conditions defined by \eqref{Eq:Q1zerocutoffboundaryconditions} then it has at least a $\mathcal{N}=1$ supersymmetry.}

{\ To get $\mathcal{N}=2$ it is necessary to ensure that $\hat{Q}_{2}$ is also self-adjoint. The procedure is no different from that which we used to determine the self-adjoint boundary conditions of $\hat{Q}_{1}$ albeit the expression for \eqref{Eq:BoundaryForm} is different (see supplementary note 2). In particular, the boundary conditions that make $\hat{Q}_{2}$ self-adjoint are
  \begin{eqnarray}
    \label{Eq:Q2zerocutoffboundaryconditions}
   \tilde{\vec{c}}^{+} = \tilde{U}_{m} \tilde{\vec{c}}^{-} \; , 
  \end{eqnarray}
where
   \begin{eqnarray}
    \tilde{c}_{j}^{\pm} &=& \frac{1}{\sqrt{2}} \left( \gamma_{j}^{\frac{1}{2}} c_{j} \pm (-i)^{N} (\gamma_{j}^{*})^{\frac{1}{2}} \tilde{c}_{N-j+1} \right) \; ,
  \end{eqnarray}
with $\tilde{U}_{m}$ a second, arbitrary $(m \times m)$-unitary matrix.}

{\ That $\hat{Q}_{1}$ and $\hat{Q}_{2}$ are self-adjoint is not enough to ensure $\mathcal{N}=2$. For this larger supersymmetry it is important that the products $\hat{Q}_{1} \hat{Q}_{2}$ and $\hat{Q}_{2} \hat{Q}_{1}$ are well defined so that \eqref{Eq:N=2SUSYdef} makes sense as an operator identity. This can only happen if both operators act on the same space of wavefunctions, which will naturally restrict $U_{m}$ and $\tilde{U}_{m}$. To determine this restriction we notice the following identity between the supercharges \cite{Falomir:2005yw}
  \begin{eqnarray}
    \label{Eq:Q+toQ-trans}
    \hat{Q}_{2} = e^{ \frac{i \pi}{4} \sigma_{3}} \hat{Q}_{1}  e^{- \frac{i \pi}{4} \sigma_{3}} \; , \qquad 
    \sigma_{3} = \left( \begin{array}{cc}
                         1 & 0 \\
                         0 & -1
                        \end{array} \right) \; . \qquad
  \end{eqnarray}
Acting on our wavefunctions with $\exp( i \pi \sigma_{3} / 4)$ allows us to translate the results for $\hat{Q}_{1}$ into results for $\hat{Q}_{2}$, and in particular relate $U_{m}$ and $\tilde{U}_{m}$. Taking some $U_{m}$ and performing the transformation one finds the following relationship
  \begin{eqnarray}
    \label{Eq:UnitaryMatrixRelation}
   \tilde{U}_{m} = i \left( U_{m} + i \mathbbm{1}_{m} \right)^{-1} \left( U_{m} - i \mathbbm{1}_{m} \right) \; ,
  \end{eqnarray}
giving $\tilde{U}_{m}$ as a function of $U_{m}$.}

{\ It follows from \eqref{Eq:UnitaryMatrixRelation} and $U_{m}$ being unitary that $\tilde{U}_{m}$ is an anti-Hermitian unitary matrix. Thus $\tilde{U}_{m}$ can only have the purely imaginary eigenvalues $\pm i$. Conversely $U_{m}$ can then only have the eigenvalues $\pm 1$. Hence we come to an important result, the self-adjoint extension parameter for a system with $\mathcal{N}=2$ SUSY must satisfy
  \begin{eqnarray}
    \label{Eq:N=2SUSYbcs}
    U_{m}^{2} = \mathbbm{1}_{m} \; \;  ( \; \mathrm{or} \; \tilde{U}_{m}^{2} = - \mathbbm{1}_{m} \; ) \; , 
  \end{eqnarray}
For other choices of $U_{m}$ it can only have $\mathcal{N}=1$ SUSY with the appropriate supercharge dependent on the space of wavefunctions upon which we act. Finally we note that the component Hamiltonians $\hat{h}$ and $\hat{h}'$, for this choice of self-adjoint boundary parameter, are also self-adjoint in their own right as discussed in supplementary note 3.}

\section{Bound state energies and scattering}

{\ For any scale invariant Hamiltonian satisfying the constraints on $\Delta_{j}$ in \eqref{Eq:PowerLawConstraints} and possessing the $\mathcal{N}=2$ boundary conditions \eqref{Eq:N=2SUSYbcs} there can be no negative energy stable bound states due to positivity of the energy
  \begin{eqnarray}
        \langle \vec{\Psi} | \hat{H} | \vec{\Psi} \rangle
   &=&  \int_{r=0}^{\infty} dr \; \left| \hat{Q}_{j} \vec{\Psi} \right|^{2} \geq 0 \;, \qquad
  \end{eqnarray}
where we have employed that boundary terms for $\hat{Q}_{j}$ vanish to arrive at this result. Importantly, at positive energies, there are $m-1 \leq N-1$ linearly independent wavefunctions that decay at infinity (see supplementary note 1). The SUSY boundary conditions \eqref{Eq:N=2SUSYbcs} impose $m$ constraints. The system is overconstrained and there can be no positive energy stable bound states. The bound state spectrum is empty with $N>1$ for exactly the same reasons it was empty at $N=1$.}

{\ In passing we note that these models lack normalisable zero modes (the condition for normalisability at $r=0$ requires that zero modes diverge at infinity). Spontaneous breaking of supersymmetry is associated with the ground state of the system not being annihilated by the supercharges \cite{junker1996supersymmetric}; but in our models there is no ground state and so what style of breaking occurs is questionable. If a hard cut-off at $r=L$ is introduced, as one would do in the renormalisation approach \cite{Camblong:2000ec,Brattan:2017yzx,2018arXiv180410213B}, then we can easily find instances of \eqref{Eq:ExpandedScaleCovariantHamiltonian} with normalisable zero modes and  thus unbroken supersymmetry.}

{\ For any member of our class of $\mathcal{N}=2$ models, as there are no stable bound states, all the physics is contained in the scattering sector. However, absence of bound states should not be taken to mean that the system is scale invariant, as an anomalously generated scale can appear in the system through resonances. This is in fact what one finds for a generic $U_{m}$ satisfying \eqref{Eq:N=2SUSYbcs}. We can see this mostly readily in one-dimension where we need not concern ourselves with evaluating the sum of the partial waves over spherical harmonics.}

{\ To calculate the phase shift we must include solutions that oscillate at infinity in our positive energy set. A general positive energy solution to the energy eigenvalue equation for $\hat{h}$, allowing for oscillatory modes and ignoring boundary conditions at $r=0$, is given by a sum 
  \begin{eqnarray}
      \label{Eq:Largerbehaviour}
      \psi(r) &=& \sum_{j=1}^{N+1} \alpha_{j} \psi_{j}(r) \; , 
  \end{eqnarray}
where $\psi_{j}(r)$ behaves as
  \begin{eqnarray}
   \sim \exp\left( - \epsilon r \exp \left( i \pi \left( \frac{1}{2} + \frac{j-1}{N} \right) \right) \right) \; , 
  \end{eqnarray}
when $r$ is large. After substituting for the $\psi_{j}(r)$, given in supplementary note 1, we can compare the small $r$ behaviour of \eqref{Eq:Largerbehaviour}  against \eqref{Eq:Nearr=0BosonExpansion} to identify the $c_{j}$,
  \begin{eqnarray}
    \label{Eq:ScatteringParameters}
    c_{j} k^{-\Delta_{j}} = \prod_{l \neq j} \Gamma\left( \frac{\Delta_{l}-\Delta_{j}}{2N} \right) \sum_{s=1}^{N+1} \alpha_{s} \left( \frac{ e^{i \phi_{s}}}{2N} \right)^{\Delta_{j}}  \; , \qquad
  \end{eqnarray}
and similarly for $\tilde{c}_{j}$ modulo an additional prefactor from matching to \eqref{Eq:Nearr=0BosonExpansion}. Once again, imposing normalisability of $\psi(r)$ at $r=0$ sets the relevant $c_{j}$ or $\tilde{c}_{j}$ to zero if $\mathrm{Re}[\Delta_{j}] \leq -1/2$ giving an expression for one of the $\alpha_{s}$ in terms of the remainder. Similarly, imposing normalisability of $\tilde{\psi}(r)$ sets $\tilde{c}_{j} = 0$ if $\mathrm{Re}[\Delta_{j+N}] \leq N-1/2$ until we have $m+1$ independent $\alpha_{s}$ remaining. Up to an overall normalisation parameter the boundary conditions of \eqref{Eq:N=2SUSYbcs} will fix these remanent $\alpha_{s}$. Finally, the one-dimensional phase shift, $\delta$, is determined in terms of the asymptotic form of the resultant positive energy wavefunction by
  \begin{eqnarray}
    \label{Eq:Defofphaseshift}
    \psi(r) \sim e^{-ikr} + e^{i \delta} e^{ikr} + \ldots
  \end{eqnarray}
and is generally a function of $k$.}

\begin{figure*}[!t]
  \centering
  \begin{subfigure}
    \centering
    \includegraphics[width=\columnwidth]{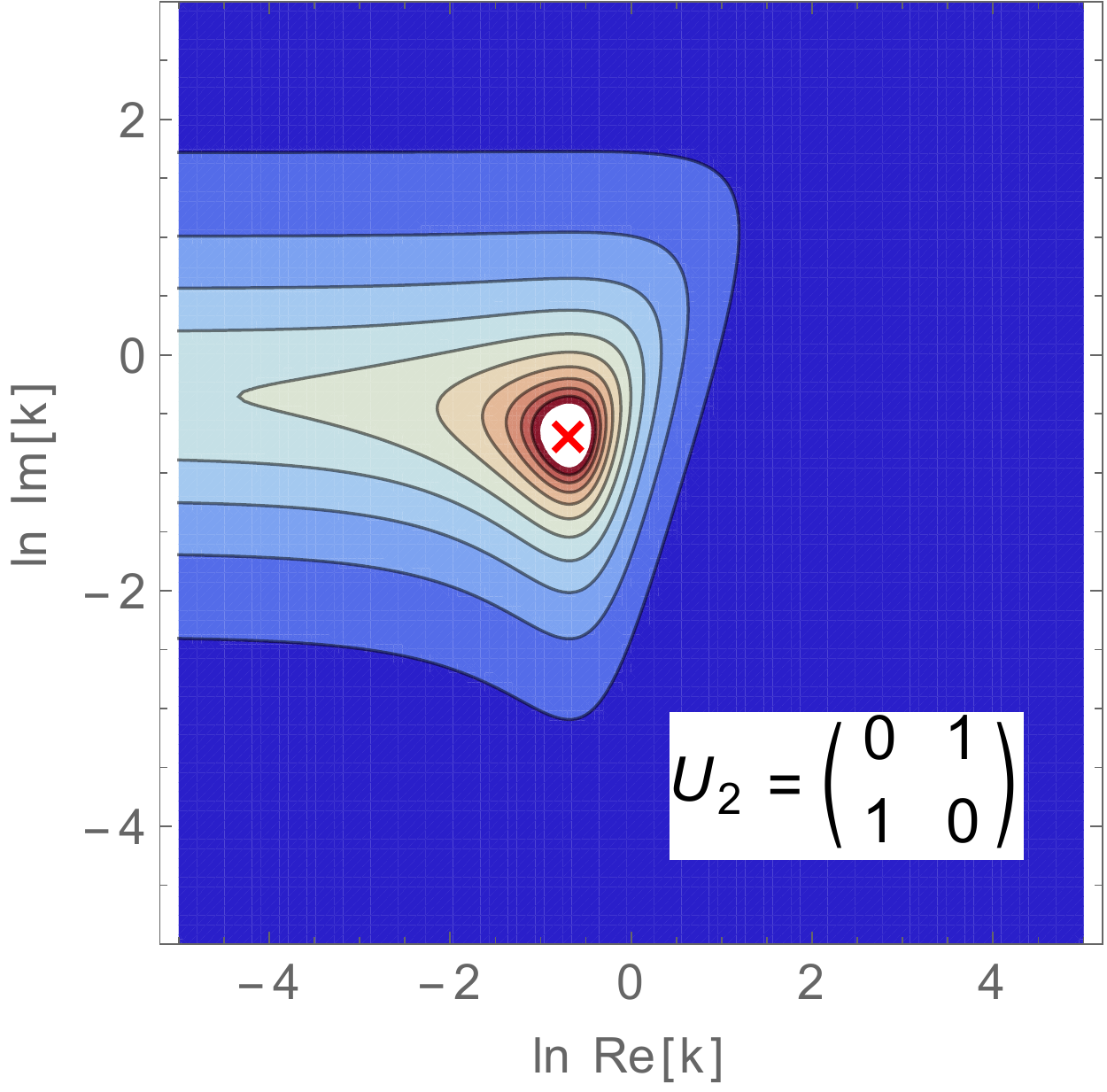}
  \end{subfigure} \hfill
  \begin{subfigure}
    \centering
    \includegraphics[width=\columnwidth]{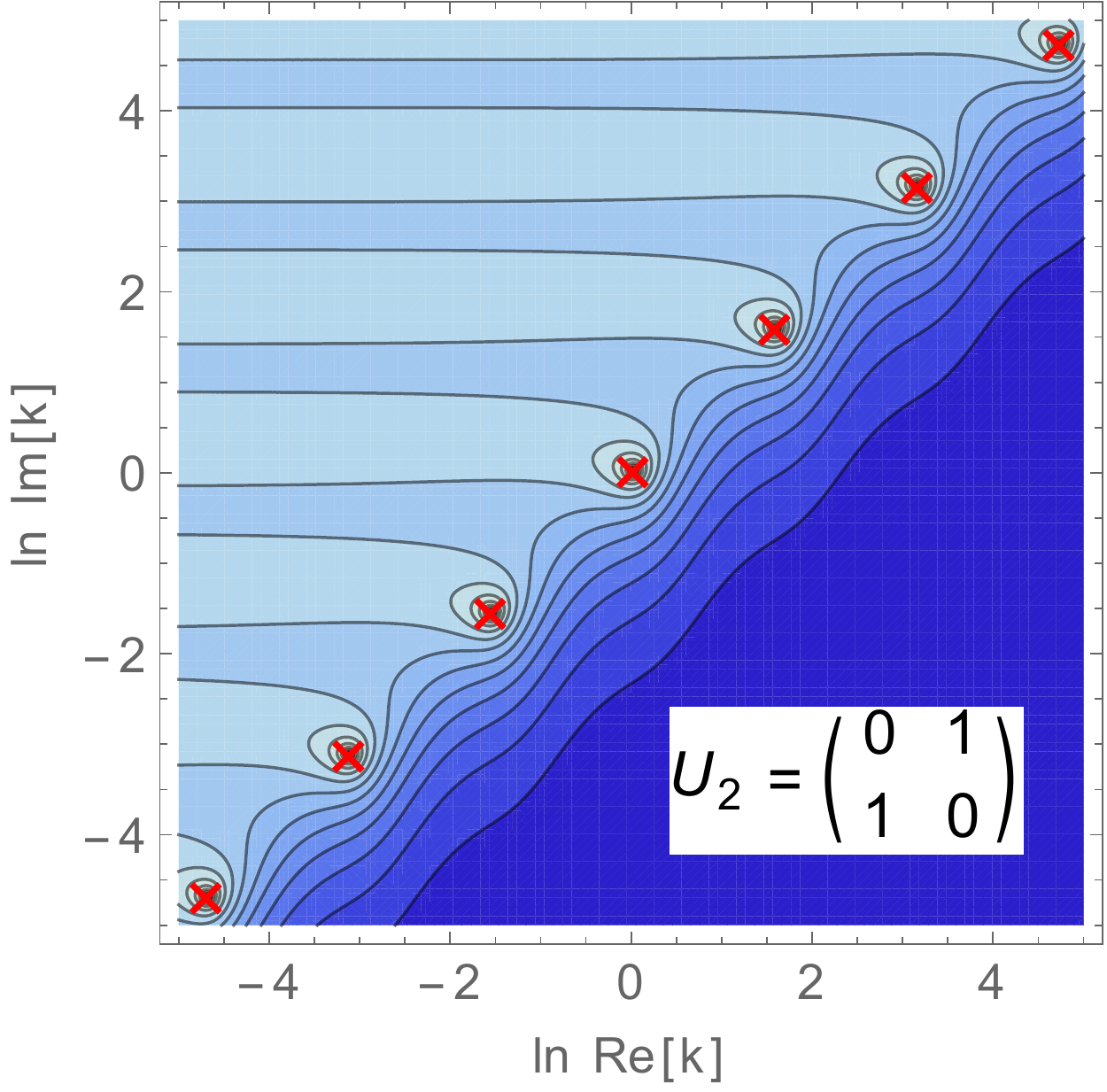}
  \end{subfigure}
  \caption{Plots of $\ln |\exp(i \delta(k))|$ with $\delta(k)$ the phase shift as a function of complex momentum for a pair of $N=2$ systems. Highlighted in the figure is the chosen self-adjoint extension parameter as substituted into \eqref{Eq:Q1zerocutoffboundaryconditions}. The displayed poles correspond to resonances of the system and thus scale invariance is broken in these cases in spite of choosing $\mathcal{N}=2$ SUSY boundary conditions. \textbf{Left:} The Hamiltonian of this system has power laws $\Delta_{1} = -1/4$ and $\Delta_{2} =1/4$ with the other two given by our constraints. In this case there is a single pole indicated by a red cross at $k \approx 0.505 (1 + i)$. \textbf{Right:} The Hamiltonian of this system has the power law $\Delta_{1} = 1/4 - i$ (the others can be determined from our constraints).  Notice that there are multiple poles, indicated by the red crosses, in the upper half of the complex plane with non-zero real momentum. They are positioned in a geometrically spaced tower at $k \approx 0.213 (1 + i) e^{\pi n/2}$ with $n \in \mathbbm{Z}$.}
  \label{Fig:N=2}
  \vspace{-1em}
\end{figure*}

{\ As an example of the appearance of resonances choose $N=2$. Unlike $N=1$, where there is essentially one form of Hamiltonian, for $N=2$ there are two types of Hamiltonian consistent with our constraints; one with completely real power laws and a second where the power laws form a complex quartet. In the first case we can parameterise the roots as $\Delta_{1}=\Delta$, $\Delta_{2}=\Delta'$, $\Delta_{3}=3-\Delta'$ and $\Delta_{4}=3 - \Delta$. Let $\Delta = -1/4$ and $\Delta'=1/4$ so that all four power laws are normalisable and choose the self-adjoint extension parameter of \eqref{Eq:Q1zerocutoffboundaryconditions} to be 
  \begin{eqnarray}
      \label{Eq:testU2}
      U_{2} = \left( \begin{array}{cc}
                   0 & 1 \\
                   1 & 0
                  \end{array} \right) \; . 
  \end{eqnarray}
We plot the $S$-matrix as a function of complex $k$ in the left hand side of fig.~\ref{Fig:N=2}. The red cross in the plot represents a pole. As this pole is at a value of $k$ with non-zero real and imaginary part it is unstable and correspondences to a resonance.}

{\ Now consider the second type of $N=2$ Hamiltonian. In this latter case we can without loss of generality parameterise $\Delta_{1}=\Delta$, $\Delta_{2}=\Delta^{*}$, $\Delta_{3}=3-\Delta$ and $\Delta_{4}=3 - \Delta^{*}$. Let $\Delta = 1/4 - i$ so that all four power laws are normalisable and choose the same self-adjoint extension parameter \eqref{Eq:testU2}. We plot the $S$-matrix as a function of complex $k$ on the right hand side of fig.~\ref{Fig:N=2}. The red crosses in the plot represent poles and form a geometric tower of quasi-bound states (not dissimilar to what is seen in \cite{ovdat2017observing}).}

{\ Despite the fact that most $\mathcal{N}=2$ SUSY self-adjoint extension parameters give resonances there is a special class of $U_{m}$ which do not (and thus the resultant model has scale invariance). This class is given by those $U_{m}$ which are diagonal matrices. To see why this should be the case we first review the special situation of $N=1$. All scale invariant $N=1$ systems with real coupling constants have the form
  \begin{eqnarray}
    \hat{h}_{S} = - \left( d_{r}^2 + \frac{\lambda}{r^2} \right) \; . 
  \end{eqnarray}
For $\lambda < 1/4$ the power laws
  \begin{eqnarray}
    \label{Eq:N=1powerlaws}
    \Delta_{\pm} = \frac{1}{2} \pm \sqrt{\frac{1}{4} - \lambda} \; ,
  \end{eqnarray}
satisfy all our constraints. Moreover if $-3/4 < \lambda < 1$ they are both real and normalisable leading to a $U(1)$ self-adjoint extension. The two $\mathcal{N}=2$ SUSY choices of boundary parameter are $U_{1} = \pm 1$ and substituting into \eqref{Eq:Q1zerocutoffboundaryconditions} we see that these choices require either $c_{1} = 0$ or $\tilde{c}_{1} = 0$. From \eqref{Eq:ScatteringParameters} and \eqref{Eq:Defofphaseshift} we subsequently determine that
  \begin{eqnarray}
    \delta_{\pm} = - \pi \Delta_{\pm} \; , 
  \end{eqnarray}
where $\Delta_{1}=\Delta_{+}$, the subscripts indicate the choices of $U_{1} =\pm 1$ respectively and $\Delta_{\pm}$ is defined in \eqref{Eq:N=1powerlaws}. As there is only one scale in the scattering problem, $k$, for a scale invariant system of one-dimension the only possible outcome was for $\delta$ to be constant; precisely as we have found. If we flip the ordering of the power laws \eqref{Eq:N=1powerlaws} the two phase shifts exchange their values.}

{\ In the case of $N=1$ the choice of SUSY parameter fixed one of the $c_{j}$ or $\tilde{c}_{j}$ to be zero, leaving the other undefined. Therefore when solving for the $\alpha_{s}$ all dependence on $k$ in \eqref{Eq:ScatteringParameters} dropped out. The previous $N=2$ case we considered did not have this property. However, if the $U_{m}$ are diagonal matrices with $\pm 1$ on the diagonal we can see from \eqref{Eq:Q1zerocutoffboundaryconditions} that this sets half of the $c_{j}$ and $\tilde{c}_{j}$ to zero, leaving the other half undefined. Hence, no momentum dependence enters into the $\alpha_{s}$.}

{\ As an example of a scale invariant theory with $N>1$ return to our previous $N=2$ system with complex power laws and substitute one of the self-adjoint extension parameters
  \begin{eqnarray}
   \label{Eq:U2scaleinvariantparam}
   U_{2} = \pm \left( \begin{array}{cc}
                   1 & 0 \\
                   0 & -1
                  \end{array} \right)
  \end{eqnarray}
into \eqref{Eq:Q1zerocutoffboundaryconditions}. Solving \eqref{Eq:ScatteringParameters} for $\alpha_{2}$ and $\alpha_{3}$ with the plus sign gives:
  \begin{eqnarray}
   \alpha_{2} = -2 \alpha_{1} e^{-\frac{i \pi}{4}} \cos\left( \frac{\pi}{2} ( \Delta^{*} - 1/2 ) \right) \; , \; \; \alpha_{3} = - i \alpha_{1} \; , \qquad
  \end{eqnarray}
with $\alpha_{1}$ a free complex parameter which will be used to normalise the wavefunction. If we choose the negative sign in \eqref{Eq:U2scaleinvariantparam} then replace $\Delta^{*}$ by $\Delta$ in the above expression. The asymptotic form of the wavefunction given the value of $\alpha_{2}$ and $\alpha_{3}$ is
  \begin{eqnarray}
   \psi(r) = \alpha_{1} \left[ \left( e^{-i k r} - i e^{i k r} \right) + \mathcal{O}(e^{-kr}) \right] \; . 
  \end{eqnarray}
The phase shift, equal to $-\pi/2$, is once again a constant independent of $k$ as required by scale invariance. If one instead chooses $U_{2} = \pm \mathbbm{1}_{2}$, the other possible choices of diagonal matrix, this phase depends on the real part of $\Delta$ i.e.~$\exp(i \delta) = \mp \exp\left( \mp \frac{i \pi}{2} \Re[\Delta] \right)$.}

{\ In conclusion, we have determined that there is a class of Hamiltonians with an anisotropic scaling symmetry that is unbroken at the quantum level. These models belong to a subclass of those with $\mathcal{N}=2$ SUSY. The supersymmetry is broken by the lack of zero modes but nonetheless constrains the energy spectrum to be absent of bound states. This ensures that all the physics of the problem is contained in the scattering sector. Moreover, for those $\mathcal{N}=2$ models that do not have scale invariance, the scaling symmetry is not broken by the presence bound states, but rather by that of quasibound states.}
  
\begin{acknowledgments}
{\ The work of DB is supported by key grants from the NSF of China with grant numbers:~11235010 and 11775212. DB would also like to thank Omrie Ovdat for comments on the draft and Mikhail Ioffe for an enlightening discussion on polynomial supersymmetry.}
\end{acknowledgments}


\bibliography{references}

\clearpage


\section*{Supplementary material}

# $\mathcal{N} = 2$ supersymmetry and anisotropic scale invariance
# Supplementary materials

Daniel K. Brattan[*]

*Interdisciplinary Center for Theoretical Study, University of Science*
*and Technology of China, 96 Jinzhai Road, Hefei, Anhui, 230026 PRC.*

(Dated: May 18, 2018)

**Supplementary Note 1 : Analytic solution of $\hat{Q}_j \Psi = \nu \Psi$**

There exist two distinct cases to consider: $\nu = 0$ and $\nu \neq 0$. For $\nu = 0$, there are $2N$ independent solutions given by $\Psi \propto r^{\Delta_i}$, $i = 1, \ldots, 2N$ obtained by inserting $\Psi \propto r^{\Delta}$ into $\hat{Q}_j \Psi = 0$ and solving for the roots of the resultant polynomials in $\Delta$:

$$\hat{Q}_j \Psi = \begin{pmatrix} (\Delta - \Delta_1) \ldots (\Delta - \Delta_N) \\ (\Delta - \Delta_{2N} + N) \ldots (\Delta - \Delta_{N+1} + N) \end{pmatrix} r^{\Delta - N}$$
$$= 0 \, .$$

For $\nu \neq 0$ the eigenvalue equation of $\hat{Q}_1$ can be rearranged into the form

$$\left( \hat{h} - \nu^2 \right) \psi(r) = 0 \, , \qquad (S1)$$

$$\tilde{\psi}(r) = \frac{1}{\nu} \hat{D}_N \ldots \hat{D}_1 \psi(r) \, . \qquad (S2)$$

Allowing $\nu$ to have an arbitrary complex value, the general solution of (S1) is expressed in terms of generalized hypergeometric functions ${}_0F_{2N-1}$ [S1]:

$$\psi(r; \phi_l) = \sum_{m=1}^{2N} e^{i \Delta_m \phi_l} \left( \frac{\epsilon r}{2N} \right)^{\Delta_m} \Gamma \left( \frac{\boldsymbol{\Delta}_m - \Delta_m}{2N} \right)$$
$${}_0F_{2N-1} \left[ \begin{array}{c} - \\ 1 - \frac{\boldsymbol{\Delta}_m - \Delta_m}{2N} \end{array}; |\nu^2| \left( \frac{r}{2N} \right)^{2N} \right] \, , \quad (S3)$$

$$\phi_l = \frac{1}{2N} (\theta - 2\pi(N + w + l)) \, ,$$
$$l = 1, 2, \ldots, 2N \, , \qquad (S4)$$

where $\boldsymbol{\Delta}_m$ is the vector of solutions to

$$(\Delta - \Delta_1) \ldots (\Delta - \Delta_{2N}) = 0$$

with $\Delta_m$ omitted, $w$ is an integer chosen such that $|\phi_l| < \pi(1 + \frac{1}{2N})$, $\nu^2 = |\nu^2| e^{i\theta}$ and $\epsilon = |\nu^2|^{\frac{1}{2N}}$. At large $r$, the wavefunctions of (S3) behave like

$$\psi(r; \phi_l) \sim \exp\left(-\epsilon r e^{i\phi_l}\right) \qquad (S5)$$

for non-zero $\epsilon$ (see [S1]). The $\psi_j(r)$ of equation (27) in the main text can be identified by examining (S5) and comparing against equation (28). We have chosen the ordering of $\psi_j(r)$ such that $\psi_1(r) \sim e^{-ikr}$ and $\psi_{N+1}(r) \sim e^{ikr}$.

To obtain the solution to $\hat{Q}_2 \Psi(r) = \nu \Psi(r)$ we use the eigenvalue equations

$$\left( \hat{h} - \nu^2 \right) \psi(r) = 0 \, , \qquad (S6)$$

$$\tilde{\psi}(r) = -\frac{i}{\nu} \hat{D}_N \ldots \hat{D}_1 \psi(r) \, . \qquad (S7)$$

Following the same procedures one finds precisely the same results.

**Supplementary Note 2 : Boundary forms**

As stated in the main text the self-adjoint extensions of $\hat{Q}_j$ are in correspondence with setting the boundary term given by evaluating

$$\int_{r=0}^{\infty} dr \left[ \boldsymbol{\Psi}^{\dagger}(r) \hat{Q}_j \boldsymbol{\Psi}(r) - \left( \hat{Q}_j \boldsymbol{\Psi}(r) \right)^{\dagger} \boldsymbol{\Psi}(r) \right] \, , \quad (S8)$$

to zero where $\boldsymbol{\Psi}(r) = (\psi(r), \tilde{\psi}(r))$. For small $r$ any solution to the eigenvalue equation for $\hat{Q}_1$, (S1) and (S2), has the form

$$\psi(r) = \sum_{j=1}^{N} c_j r^{\Delta_j} \left(1 + \mathcal{O}(r^{2N})\right)$$
$$+ (-i)^N \nu \sum_{j=1}^{N} \alpha_j \tilde{c}_j r^{\Delta_{N+j}} \left(1 + \mathcal{O}(r^{2N})\right) \, , \quad (S9)$$

and

$$\tilde{\psi}(r) = \sum_{j=1}^{N} \tilde{c}_j r^{\Delta_{N+j} - N} \left(1 + \mathcal{O}(r^{2N})\right)$$
$$+ (-i)^N \nu \sum_{j=1}^{N} \beta_j c_j r^{\Delta_j + N} \left(1 + \mathcal{O}(r^{2N})\right) \, , \quad (S10)$$

where

$$(\alpha_j)^{-1} = \prod_{l=1}^{N} (\Delta_{j+N} - \Delta_l) \, , \qquad (S11)$$

$$(\beta_j)^{-1} = \prod_{l=1}^{N} (\Delta_j - \Delta_{l+N} + 2N) \, . \qquad (S12)$$

---

[*] danny.brattan@gmail.com

Substituting these solutions for $\psi(r)$ and $\tilde{\psi}(r)$ into (S8) we find

$$i^N \sum_j \left( \gamma_j c_j \tilde{c}^*_{N-j+1} - (-1)^N \gamma_j^* c_j^* \tilde{c}_{N-j+1} \right) + \mathcal{O}(r^{2N}) \,,$$

where $j$ runs only over values with $\text{Re}[\Delta_j] > -1/2$ and $\text{Re}[\Delta_{N+j}] > N + 1/2$. The numbers $\gamma_j$ are complicated functions of the $\Delta_l$. As promised, it is a sesquilinear form in $c_j$ and $\tilde{c}_j$. Similarly, for $\hat{Q}_2$ one finds

$$i^{N+1} \sum_j \left( \gamma_j c_j \tilde{c}^*_{N-j+1} - (-1)^{N+1} \gamma_j^* c_j^* \tilde{c}_{N-j+1} \right)$$
$$+ \mathcal{O}(r^{2N}) \,.$$

**Supplementary Note 3 : Self-adjointness of $\hat{h}$ and $\hat{h}'$**

Self-adjoint extensions of $\hat{h}$ are in correspondence with boundary conditions that set

$$\int_{r=0}^{\infty} dr \, \left[ \phi(r) \hat{h} \psi(r) - \left( \hat{h} \phi(r) \right)^\dagger \psi(r) \right] \,, \quad (S13)$$

to zero non-trivially. Assuming decay at infinity this integral evaluates to a boundary term. Similarly for $\hat{h}'$. Denoting the boundary forms of $\hat{H}$, $\hat{Q}_j$, $\hat{h}$ and $\hat{h}'$ by $\Delta$ with the operator as a subscript it can be shown at the level of the integrals that

$$\Delta_H[\mathbf{\Phi}, \mathbf{\Psi}] = \Delta_{Q_j}[\hat{Q}_j \mathbf{\Phi}, \mathbf{\Psi}] + \Delta_{Q_j}[\mathbf{\Phi}, \hat{Q}_j \mathbf{\Psi}]$$
$$= \Delta_h[\phi, \psi] + \Delta_{h'}[\tilde{\phi}, \tilde{\psi}] \,.$$

For the case of $\hat{h}$ the small $r$ expansion of every wavefunction has the form

$$\psi(r) = \sum_{j=1}^N d_j r^{\Delta_j} \left( 1 + \mathcal{O}(r^{2N}) \right)$$
$$+ \sum_{j=1}^N \tilde{d}_j r^{\Delta_j + N} \left( 1 + \mathcal{O}(r^{2N}) \right) \quad (S14)$$

and the corresponding boundary form is

$$\sum_j \left( \epsilon_j d_j \tilde{d}^*_{N-j+1} - \epsilon_j^* d_j^* \tilde{d}_{N-j+1} \right) \,, \quad (S15)$$

where $\epsilon_j$ is some number dependent on the $\Delta_l$. In writing (S15) we have implicitly set to zero any $d_j$ or $\tilde{d}_{N-j+1}$ whose associated power law has a real part less than $-1/2$. The coefficients $d_j$ and $\tilde{d}_{N-j+1}$ should be compared to $c_j$ and $\tilde{c}_{N-j+1}$ of the generic, small $r$, supersymmetric solutions in the main body of the paper. Assume for now that $\text{Re}[\Delta_{N+j}] > N - 1/2$ for all $j$. The boundary conditions that make $\hat{h}$ self-adjoint are

$$\boldsymbol{d}^+ = V \boldsymbol{d}^- \,, \quad (S16)$$

where

$$d_j^\pm = \frac{1}{\sqrt{2}} \left( \epsilon_j^{1/2} d_j \pm i (\epsilon_j^*)^{1/2} \tilde{d}_{N-j+1} \right) \,, \quad (S17)$$

and $V$ is an arbitrary $(m \times m)$-unitary matrix. We choose a convention on the roots $(\epsilon_j)^{1/2}$ and $(\epsilon_j^*)^{1/2}$ such that $(\epsilon_j^*)^{1/2} = \left( \epsilon_j^{1/2} \right)^*$.

We can rearrange (S16) into the form

$$\left( \mathbb{1}_m - \Gamma V \Gamma^\dagger \right) \boldsymbol{d} = \Gamma^{-1} \Gamma^\dagger \left( \mathbb{1}_m + \Gamma V \Gamma^\dagger \right) \tilde{\boldsymbol{d}} \,, \quad (S18)$$
$$\Gamma_{jj} = (\epsilon_j)^{\frac{1}{2}} \,, \quad (S19)$$

where $\boldsymbol{d}$ and $\tilde{\boldsymbol{d}}$ are the vectors whose $j^\text{th}$ components are $d_j$ and $i\tilde{d}_{N-j+1}$ respectively. From the main body of the paper we know that the supersymmetric boundary conditions are such that $U_m^2 = \mathbb{1}_m$. This allows us to rewrite the SUSY boundary conditions as

$$(\mathbb{1}_m - U_m) \boldsymbol{c} = 0 \,, \quad (S20)$$
$$(\mathbb{1}_m + U_m) \tilde{\boldsymbol{c}} = 0 \,, \quad (S21)$$

where we have used the fact that $\frac{1}{2}(\mathbb{1}_m \pm U_m)$ are mutually orthogonal projectors. Here the $j^\text{th}$ components of $\boldsymbol{c}$ and $\tilde{\boldsymbol{c}}$ are $c_j$ and $\tilde{c}_{N-j+1}$ respectively. We see then that if we identify $U_m = \Gamma V \Gamma^\dagger$ we have interpreted the self-adjoint extension of the supersymmetric system in terms of a self-adjoint extension of $\hat{h}$. This allows us to perform calculations with $\hat{h}$, such as determining the phase shift, without considering the full SUSY system.

In the above calculation we assumed that $\text{Re}[\Delta_{N+j}] > N - 1/2$ for all $j$ which resulted in $U_m$ and $V$ being of the same dimension. However, if there was such a power law in the generic expansion (S14) that satisfied $-1/2 \leq \Delta_{N+j} \leq N - 1/2$ then this power law must be included in the self-adjoint extensions of $\hat{h}$ but not in the self-adjoint extension of $\hat{Q}_j$. To match the SUSY result we must choose a self-adjoint extension of $\hat{h}$, i.e. a $V$, such that the corresponding coefficient of this power law $\tilde{d}_{N-j+1}$ in (S15) is identically to zero. Assuming that this occurs for $j = 1$, the corresponding self-adjoint extension is then

$$\begin{pmatrix} -1 & 0 \\ 0 & U_m \end{pmatrix} = \Gamma V \Gamma^\dagger \quad (S22)$$

where now $V$ and $\Gamma$ are $(m+1) \times (m+1)$ matrices. A similar situation can occur for the self-adjoint extension of $\hat{h}'$ and the remedy is no different, we augment the SUSY parameter $U_m$ with $\pm 1$ until it is of the appropriate size.



\end{document}